\documentclass{article}
\usepackage[english]{babel}
\usepackage[margin=1in]{geometry}
\usepackage{multicol}
\usepackage{background}
\usepackage[math]{blindtext}% just to generate filler text

\usepackage{epstopdf}
\begin{document}
\bibliographystyle{unsrt}

\title{{\bf Impact of giant resonant dispersion on the response of intracavity} \\
{\bf phase interferometry and laser gyros}}
\author{James Hendrie$^1$, Matthias Lenzner$^2$,  Jean Claude Diels$^1$ and Ladan Arissian$^{1,*}$\\ \\ \small
$^1$ CHTM, University of New Mexico, Albuquerque, New Mexico 87106 \\
\small $^2$ Lenzner Research, LLC, 125 E canyon View dr,  Tucson, AZ 85704 \\
\small email $^*$ ladan@unm.edu}
\normalsize

%\dates{}
\maketitle
%\ociscodes{(140.4050)   Mode-locked lasers;  (120.5050)    Phase measurement; (120.2230) Fabry-Perot;
 % (320.7090) Ultrafast lasers.}

\begin{abstract}
Intracavity Phase Interferometry is a phase sensing technique using mode-locked lasers
in which two intracavity pulses circulate. The beat frequency between the two output frequency combs
is proportional to a phase shift to be measured.  A laser gyro is a particular implementation of this device.
The demonstrated sensitivity of $10^{-8}$ could be manipulated by applying
a giant dispersion to each tooth of the comb.  Such coupling is achieved
with an intracavity etalon, resulting  a
 large change in phase response
of a ring  laser.  This change is shown to be unrelated to the average
pulse velocity within the laser cavity.

\end{abstract}

\section{Introduction}

There have been numerous studies of ``slow light'' and ``fast light'' where
resonant structures~\cite{Smith08} have been used to enhance or
reduce the slope of the dispersion $dk/d\Omega$, where $k(\Omega)$ is the wave vector as a function of light angular
frequency $\Omega$.  One application that has been
proposed is to increase the Sagnac response, hence
create passive~\cite{Leonardt00} or active~\cite{Shahriar07} optical gyroscopes of enhanced sensitivity.
These studies associate the enhancement or decrease of the gyro response with
a change in pulse velocity.  Even though these proposals were dealing with group velocities, they were aimed exclusively at
cw lasers, where there is no propagating wave packet, and it is difficult to associate the claimed enhancement/reduction of a
gyroscopic (Sagnac) response  with a change in group velocity.
Furthermore, as has been pointed out by Arnold Sommerfeld in 1907 already~\cite{Sommerfeld1907},
the mathematical quantity group velocity does not represent the velocity of an
electromagnetic signal in frequency regions with large dispersion~\cite{BrillouinBook}.

Measurements presented here of the gyro response in a mode-locked
ring laser as a function of average pulse velocity in the cavity
demonstrate that the
concept of ``fast light'' or ``slow light'' applied to this situation is a misnomer since:
   \begin{itemize}
\item  Modifications of the
   average pulse velocity of the laser leaves the gyro response unaffected.
\item The gyro response can indeed be altered by the dispersion of an intracavity element.
However, the repetition rate of the laser is not solely determined by this dispersion.
\end{itemize}
The expression ``gyro response'' is usually understood as  the beat note frequency measured
between the two outputs corresponding to oppositely circulating beams in a ring laser, as a
function of the rotation of the laser about an axis orthogonal to its plane.  This is only a particular case
of ``Intracavity Phase Interferometry'' (IPI) where a mode-locked laser is used as an
interferometer in which two pulses circulate independently~\cite{Arissian14b}.  The frequency of the beat note
obtained by interfering
the two output frequency combs (corresponding to each of these intracavity pulses)
is a measure of a differential phase shift at each round trip between the circulating pulses.
IPI in this paper refers to the phase interferometry inside a {\em laser cavity} as opposed to the
enhancement of the phase response associated with a {\em passive Fabry-Perot}~\cite{Ma99}.
While the distinction may appear merely quantitative because the laser can be seen as a
Fabry-Perot of extreme finesse, the difference is qualitative because the response is
an optical frequency shift rather than an amplitude modulation.
Another distinction has to be made with the label
``active-cavity interferometer'' as used by Abramovici and Vager~\cite{Abramovici85}, where
two gain media are inserted in 2 branches of a Michelson,
which result in uncorrelated spontaneous emission noise.  In IPI, there is only one gain medium
acting on both pulses.

The average velocity of a pulse circulating inside
a mode-locked laser is
continuously tuned by adjusting the angle of incidence of a Fabry-Perot etalon inserted
in the laser cavity~\cite{Masuda16}.  By applying this technique to a ring laser,
it is demonstrated here that
the gyro response is not correlated to the
pulse envelope velocity.  However, because the Fabry-Perot is coupled to the mode-locked laser cavity,
 the resonant {\em dispersion} of the etalon creates a significant change in
the phase (or gyro) response of the laser.

\section{Phase response in a mode-locked laser}
\label{phase_response}
In the general case of ``Intracavity Phase Interferometry'' (IPI), which involves linear as well
as ring mode-locked lasers,  a physical quantity to be measured
 (nonlinear index, magnetic field, rotation, acceleration, electro-optic coefficient,
 fluid velocity, linear index) creates a differential phase shift $\Delta \phi$ between the two pulses, which,
 because of the resonance condition of the laser, is translated into a difference in optical
 frequency~\cite{Arissian14b}.  This difference  is measured as a beat note
 produced when interfering the two frequency combs generated by the laser.  The measured beat note
$\Delta \omega$ can be expressed as:
 \begin{equation}
 \Delta \omega = \frac{\Delta \phi}{\tau_\phi} = \omega \frac{\Delta P}{P},
 \label{basic_phase_response}
 \end{equation}
where $\tau_\phi$ is the round-trip time of the pulse circulating in a  laser cavity of perimeter $P$ (in the
case of a linear cavity of length $L$, $P = 2L$ and $\Delta P = 2 \Delta L$), and $\omega$ is the average
optical pulse frequency. The technique of IPI has been shown to have extreme sensitivity, with the
ability to resolve phase shift differences as small as $\Delta \Phi \approx 10^{-8}$
(corresponding to a beat note bandwidth of 0.16 Hz for a cavity of $\tau_\phi$ =
10 ns~\cite{Arissian14b,Velten10}).
This corresponds to an optical path difference of only 0.4 fm.  If applied to a square ring laser
of 4 m$^2$, the beat note bandwidth of 0.16 Hz corresponds to a sensitivity in
rotation rate change of $\approx 0.2$ revolution/year.

The principle of the ``fast light enhancement'' of the response of intracavity phase interferometry
(and in particular gyro response) is to make $\tau_\phi$ frequency dependent through
an element having a transfer function
$\tilde{\cal T}(\Omega) = \left | \tilde{\cal T}\right | \exp[-i \psi(\Omega)]$ with giant dispersion:
\begin{equation}
\tau_\phi = \tau_{\phi 0} +  \left . \frac{d\psi}{d\Omega} \right |_{\omega_0}
\label{dispersion}
\end{equation}
where $\tau_{\phi 0} =   (P n_p)/c$ is the round-trip time without dispersive element, $n_p$ is the {\em phase}
index of refraction  at the central carrier frequency ${\omega_0}$ averaged over the elements of the cavity.
$-\psi(\Omega)$ is the phase of the transfer function of the dispersive  optical element inserted in the cavity, with $\psi(\omega) = 0$.
By substituting Eq.~(\ref{dispersion}) in Eq.~(\ref{basic_phase_response}), the beat note is thus:
\begin{equation}
\Delta \omega = \frac{\frac{d \phi}{\tau_{\phi 0}}}{1 +
\frac{1}{\tau_{\phi 0}}\left .\frac{d \psi}{d\Omega} \right |_{\omega_0}} =
\frac{\Delta \omega_0}{1 +
\frac{1}{\tau_{\phi 0}}\left .\frac{d \psi}{d\Omega} \right |_{\omega_0}}
\label{basic_equation_disp}
\end{equation}
It should be noted that all the above considerations pertain to {\em phase} resonances and
velocities.  In the case of normal dispersion, $d\psi/d\Omega|_{\omega_0}$ is positive,
resulting in a decrease of $\Delta \omega$.   There is amplification of the phase response
if $d\psi/d\Omega|_{\omega_0}$ is negative, a case that
is most often quoted as a ``fast light'' response.
If we consider simply propagation through a transparent medium, $\psi = [k(\Omega)-k_0] d$,
where $k(\Omega) = \Omega n(\Omega)/c$,  is the wavevector of a medium of
thickness $d$ and index $n(\Omega)$, and $k_0 = k(\omega_0)$, then
the second term in the denominator of Eq.~(\ref{basic_equation_disp}) is:
\begin{equation}
\frac{1}{\tau_{\phi 0}}\left .\frac{d \psi}{d\Omega} \right |_{\omega_0} = \frac{1}{\tau_{\phi 0}}\frac{d}{v_g},
\label{group-phase}
\end{equation}
where  $v_g$ is the group velocity {\em in a dielectric}.  Equation~(\ref{group-phase})
deals fundamentally with the {\em phase} of the light in a laser cavity, and not the envelope velocity of a
circulating pulse.  As demonstrated in reference~\cite{Masuda16}, the envelope velocity
of a pulse circulating in
a mode-locked laser is not related to $\left .dk/d\Omega \right |_{\omega_0}$ for
a $k$ vector averaged in the cavity, but to the gain and loss dynamics inside the
laser.  This point will be further emphasized in the present paper,
where it is shown that the envelope velocity of circulating
pulses or bunches of pulses can be varied, while the gyro response remains unchanged.
However, it will be shown that the teeth of the frequency comb of mode-laser
can be coupled to the modes of an intracavity etalon.
It is further demonstrated that a large dispersion  results from
this coupling, with a magnitude such that $\frac{1}{\tau_{\phi 0}}\left .\frac{d \psi}{d\Omega} \right |_{\omega_0} $
is of the order of unity.

\section{Challenge in achieving laser dispersion}
\label{challenge}

\begin{figure} [h!]
\centering
\includegraphics*[width=\linewidth]
{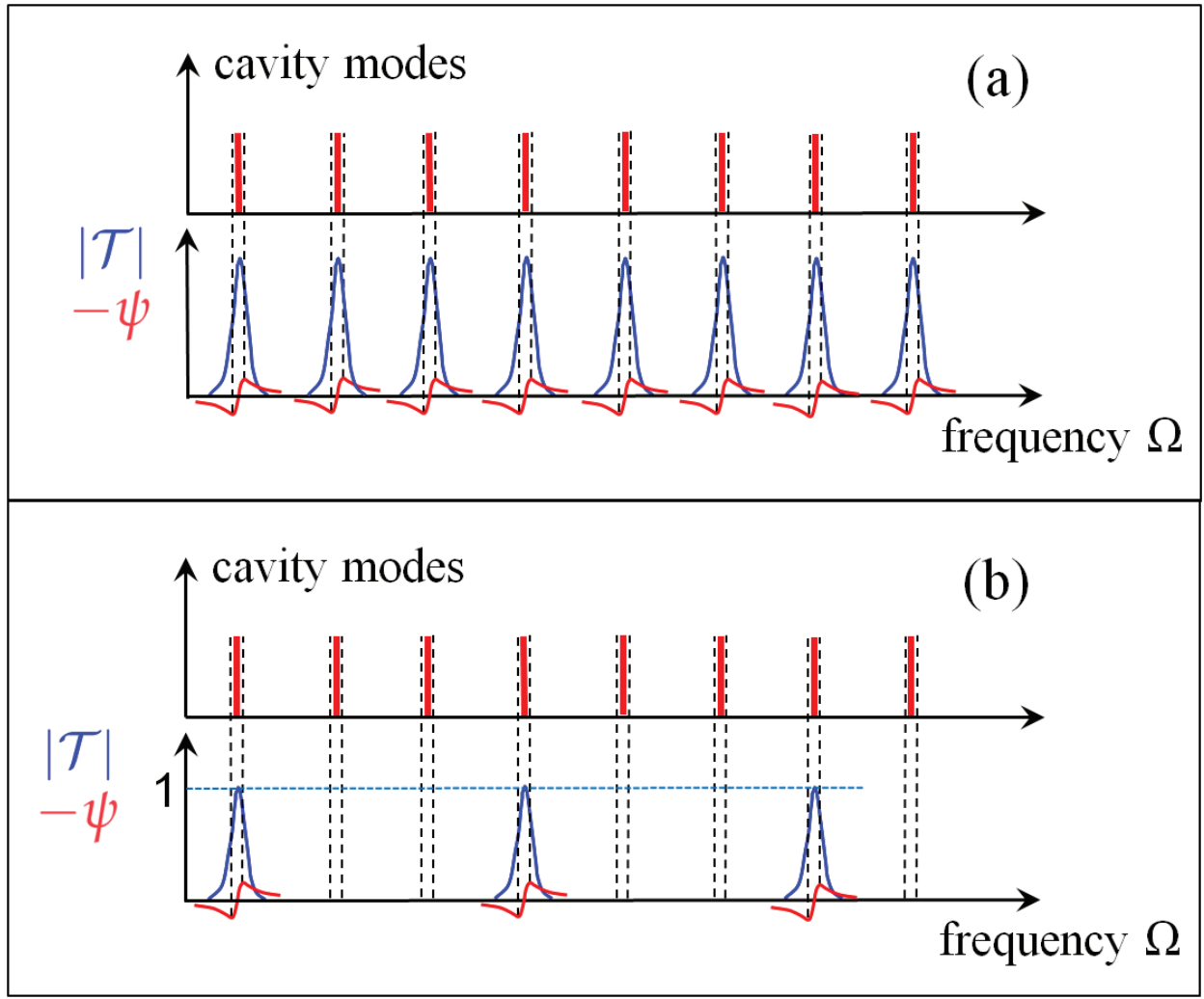} \caption[]{\small (a) Frequency comb(s) out of a ring mode-locked laser.
At rest, the teeth of the two countercirculating combs coincide (vertical red lines).
In presence of a relative phase shift/round-trip (Sagnac effect in the case of a laser gyro),
the teeth split (dashed lines).  According to Eq.~(\ref{basic_equation_disp}), this frequency splitting can be
modified by the dispersion of a sharp resonance, which should be present at each tooth of the comb at rest.
(b) If a Fabry-Perot etalon (i.e. a 15 mm thick piece of fused silica) is inserted in the laser cavity,
it has been shown~\cite{Masuda16} that each mode of the Fabry-Perot couples to a mode of the cavity.
The laser comb being locked to the Fabry-Perot modes, each tooth (or pair of teeth) experiences
the dispersion of the etalon having acquired the finesse of the laser.}
\label{teeth-modes}
\end{figure}

In order to achieve the very large dispersion required to modify the phase response through
Eq.~(\ref{basic_equation_disp}), a very narrow-band resonant structure is required.  Narrow bandwidth
implies long pulses or cw radiation, where most of the research in this field has focused.
For instance, theoretical estimates have found that large $d\psi/d\Omega$ can be produced
 by two-peak gain and coupled
resonators~\cite{Yum10,Smith14},  or by an atomic medium~\cite{Smith09}.  The latter
property has been verified experimentally.  These regions of large dispersion have a small bandwidth,
which needs only to exceed the
largest beat note to be measured. For the mode-locked laser however,
the giant slope of the resonant phase $\psi(\Omega)$ versus frequency has to be seen
by every tooth of the comb, as illustrated in Fig.~\ref{teeth-modes}.
In a mode-locked laser gyro, as with any implementation of intracavity phase interferometry, the
two circulating pulses have to meet at the same point at every round-trip~\cite{Arissian14b}. As the pulses circulating
in opposite direction see an optical length differential, decreased or augmented by the giant dispersion, one would
expect that the crossing point cannot be maintained, if the pulse velocity were simply equal to
$1/(dk/d\Omega)$. However, it has been established that the average
envelope velocity in a mode-locked laser
is dominated by gain and loss dynamics of the entire cavity,
and that the crossing point of the two pulses can be maintained~\cite{Arissian14b,Masuda16}.

\subsection{Giant dispersion of an intracavity etalon}
In order for the ``fast light enhancement'' or ``slow light reduction'' of Section~\ref{challenge} to apply to a mode-locked laser, there
should be a resonant structure for each mode of the laser, as sketched in Fig. 1 (a). The dashed lines
indicate the position of the counter-circulating modes after rotation.  The
alternative to having a resonant structure for each mode  is to have a resonant structure with much larger mode spacing
(for instance 100 $\times$ larger) locked to the modes of the laser [Fig.~\ref{teeth-modes} (b)]. The mode-locked laser is known to
create a frequency comb with equally spaced modes~\cite{Udem99a, Udem99b,Jones00}. By the same mechanism explained in
reference~\cite{Arissian09}, the teeth of the frequency comb will be locked by the modes that are resonant with
those of the resonant structure in Fig.~\ref{teeth-modes} (b). A resonant structure could be a Fabry-Perot etalon
inserted in the cavity. At first sight this seems to be an impossibility, because:
\begin{itemize}
\item If the Fabry-Perot has a high finesse, it will filter the frequency spectrum of the laser, resulting
in long pulse or cw operation
\item If the etalon has a low finesse we do not have resonant enhancement
\item A complicated electronic feedback system would be required to maintain the modes of the
laser cavity and those of the etalon locked, such as has been realized in reference~\cite{Delfyett06}.
\end{itemize}
A study made of the characteristics of a mode-locked laser with an intracavity etalon~\cite{Masuda16} contradicts all
these points. A low finesse Fabry-Perot inserted in the mode-locked laser creates a high frequency
pulse train passively locked to the comb (modes) of the laser. The low finesse uncoated etalon,
when inserted in the mode-locked cavity, acquires a high finesse determined by the laser cavity.
All these points studied with a linear laser are confirmed in the ring configuration presented in the
next section.

\section{Experiments with a mode-locked ring Ti:sapphire laser}
\begin{figure} [h!]
\centering
\includegraphics*[width=\linewidth]
{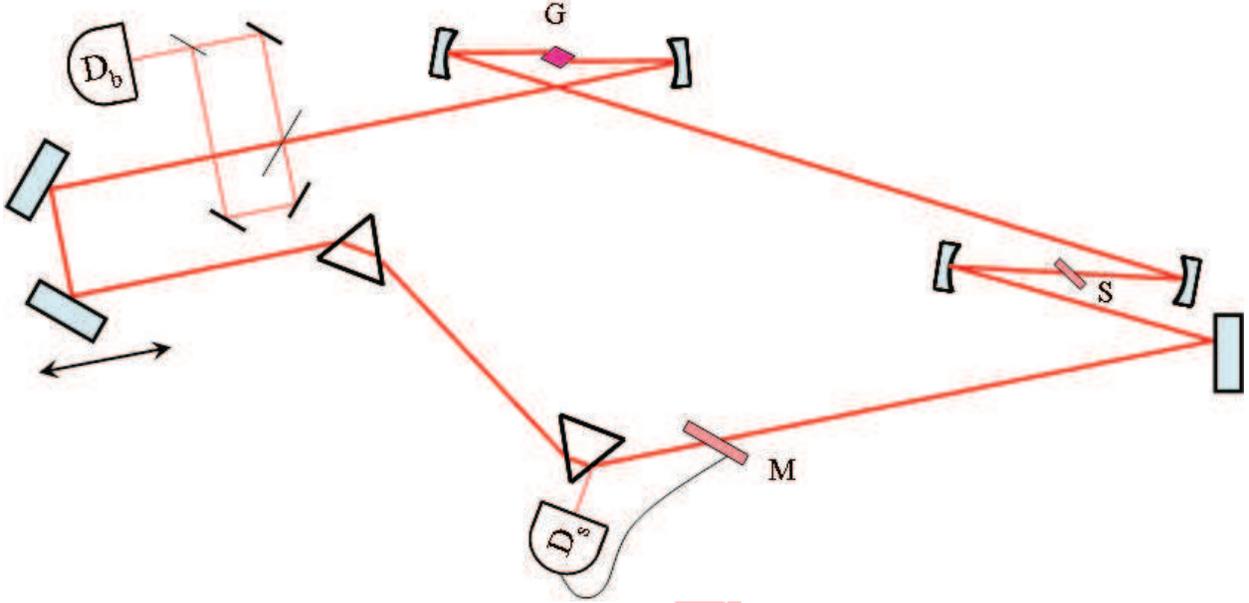} \caption[]{\small Ti:sapphire ring laser, mode-locked by the saturable absorber
Hexaindotricarbocyanine iodide (HITCI) dissolved in a jet of ethylene glycol ($S$).  The gain medium ($G$)
and the phase modulator ($M$) are located at approximately 1/4 cavity perimeter from the
saturable absorber $S$.  An output coupling is made near the other pulse crossing point, and the two
output pulse trains are made to interfere on a detector $D_b$ to monitor
the beat note between the two frequency combs.
Instead of rotating the laser, a phase shift/round-trip is provided by a phase modulator $M$ driven  by
the detector $D_s$ at the cavity repetition rate (details in reference~\cite{Arissian14b}).}
\label{ring_laser}
\end{figure}
A Ti:sapphire ring laser mode-locked by a saturable absorber was constructed for a demonstration experiment.
As sketched in Fig.~(\ref{ring_laser}), the main components of the laser are a Ti:sapphire gain crystal pumped by
a frequency doubled vanadate laser, prisms for dispersion compensation, and a saturable absorber dye jet for mode-locking
and for defining the crossing point of two pulses circulating in the cavity.  Such a ring laser  is one of the
numerous mode-locked laser gyro and IPI configurations that has been demonstrated in past
research~\cite{Arissian14b,Lai92c}.
It has also been established that, when a saturable absorber is used, it should be
in a flowing configuration (dye jet)
 to prevent phase coupling between the two counter-circulating pulses, by randomizing the phase of the backscattering of one pulse
 into the other~\cite{Arissian14b}.

The response of intracavity phase interferometry is investigated by applying a differential phase shift per round trip
with a  phase modulator inserted in the cavity, located preferably at 1/4 perimeter away from the pulse crossing point.
The phase modulator is a 100 $\mu$m thick plate of lithium niobate, oriented at Brewster angle,
with electrodes on one face to apply an electric field along the $z$ crystallographic axis.
The applied field is achieved by narrow band amplification of the signal of a detector monitoring
one of the output pulse trains (detector $D_s$ monitoring the pulse train from the clockwise circulating pulse
in Fig.~\ref{ring_laser}).

\section{Modifications of the phase response and envelope velocities with an intracavity Fabry-Perot etalon}

\subsection{Envelope velocities}

\begin{figure} [h!]
\centering
\includegraphics*[width=\linewidth]
{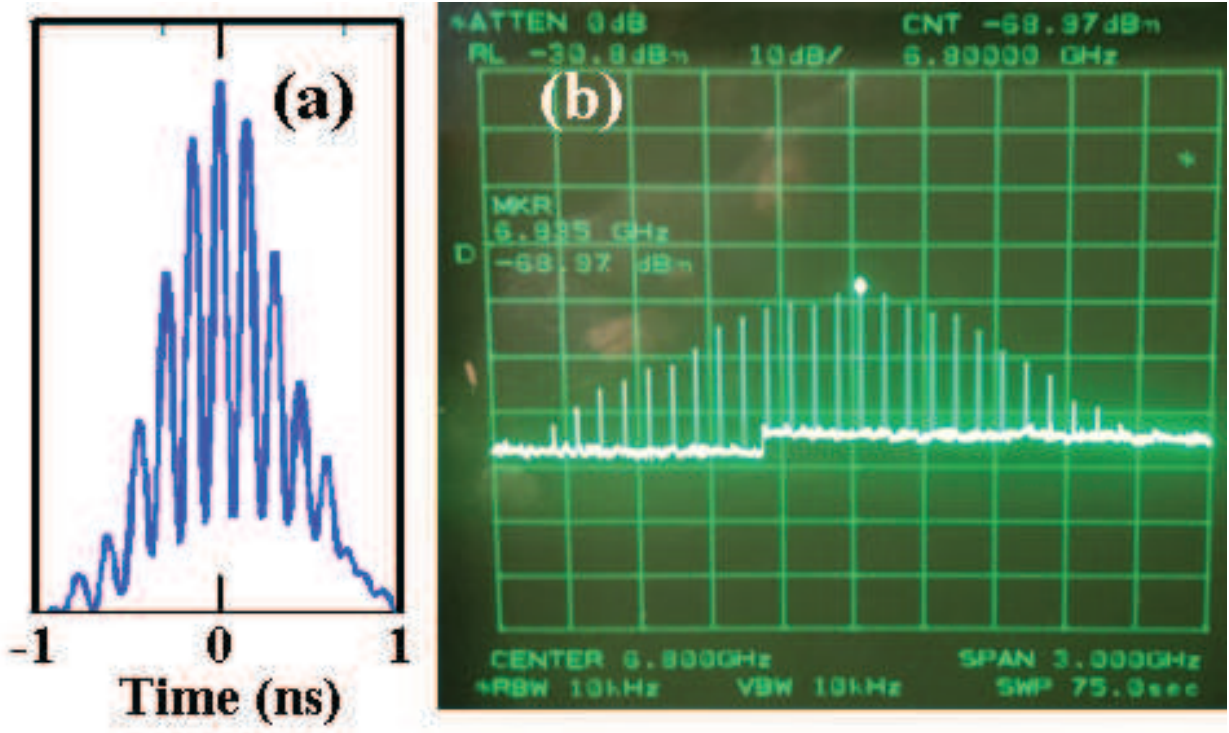} \caption[]{\small (a) oscilloscope trace of a high repetition pulse train
created by the intracavity etalon.  This picture is recorded with a fast
photodiode and an 8 GHz oscilloscope.  (b) Spectrum of the nested frequency comb,
 recorded with the same photodiode
and a spectrum analyzer.
The center frequency is 6.8 GHz, and the span 3 GHz.  The marker is at 6.835 GHz.
The baseline step is an artifact of the spectrum analyzer.}
\label{pulsetrain}
\end{figure}
A 15.119 mm thick fused silica etalon, uncoated, is inserted in the ring cavity.
This material has a phase index of $n$ = 1.4534 at the laser wavelength of 800 nm,
and a group index $n_g$ = 1.46.   Similar to the situation in a linear
mode-locked laser~\cite{Liu05,Masuda16}, despite the very low finesse of this Fabry-Perot, it influences the mode-locking
by creating a high frequency (close to the etalon round-trip frequency) pulse train
which repeats itself at a lower frequency (close to the original laser repetition rate).
Figure~\ref{pulsetrain} (a) shows the high repetition rate pulse train nested inside
the laser pulse train, which  can be explained as follows.
At every round-trip, each pulse of the high frequency train adds coherently to the next one.
This coherence is established
through the resonance condition of the pulses within the laser cavity. Figure~\ref{pulsetrain} (b) shows the
spectrum of the nested frequency comb
associated with the double pulse train.

The changes in repetition rate of the laser cavity, after introduction of the Fabry-Perot, cannot be explained by
the traditional group delay introduced by the etalon.  The transmission function of a Fabry-Perot  of thickness $d$ and intensity reflectivity $R= |r|^2$
(where $r$ is the field reflectivity), at an internal angle
$\theta$ with the normal,  is:
\begin{equation}
{\cal T}(\Omega) = \frac{(1 - R)e^{-ikd\cos \theta}}{1 - R e^{-2ikd\cos \theta}}.
\label{FPT}
\end{equation}
The group delay is the first derivative of the phase $\psi$ of this expression with respect to
frequency:
\begin{equation}
\left |\frac{d \psi}{d \Omega} \right |_{\omega_0} = \left (\frac{1 + R}{1 - R}\right )
\frac{1 + \tan^2 \delta}{\left [1 + \left( \frac{1 + R}{1 - R} \right)^2 \tan^2 \delta\right ] }\frac{nd}{c}
\label{FPgroup_delay}
\end{equation}
where $\delta = kd = \omega n d/c$.
This expression being correct near a resonance, we will make the approximation $\tan \delta \approx \delta$.  To remain within the bandwidth of the
Fabry-Perot transmission,
$(1+R)/(1-R)\delta << 1$, and:
\begin{equation}
\left |\frac{d \psi}{d \Omega} \right |_{\omega_0} \approx \frac{1+R}{1-R} \frac{nd}{c}
\label{approx1}
\end{equation}

It has been demonstrated that the average velocity of the pulse circulating in the mode-locked cavity
differ considerably from the group delay of Eq.~(\ref{FPgroup_delay}) and~(\ref{approx1}).  The group delays are determined by
dynamic gain and loss considerations.  For instance, the continuous transfer of energy from each pulse of the
high frequency train into the next one results in a delay of the center of gravity of that train.
Saturable gain has the opposite effect of accelerating the pulse trains.
The average velocities, as modified by the Fabry-Perot etalon, as function of the tilt of the etalon,
have been measured and matched with theoretical simulations in reference~\cite{Masuda16}.

\begin{figure} [h!]
\centering
\includegraphics*[width=\linewidth]
{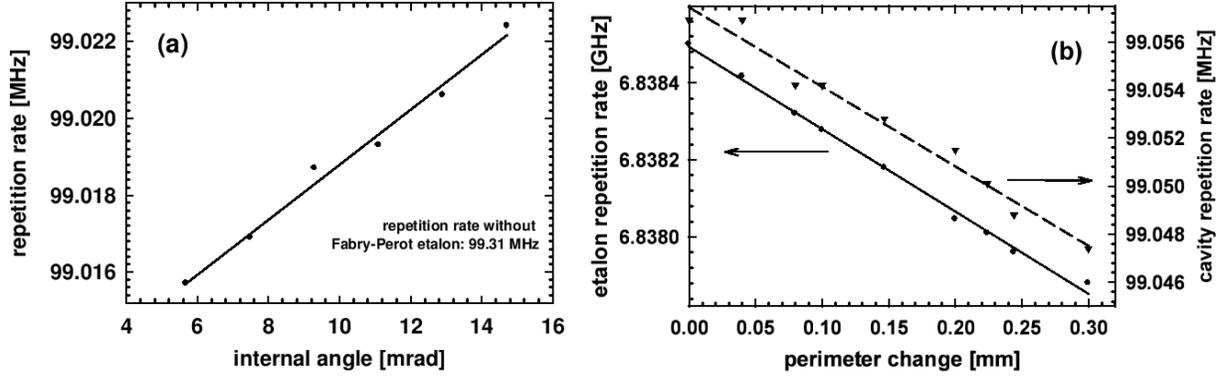} \caption[]{\small (a) Plot of the tooth spacing
of the frequency comb corresponding to the clockwise circulating
group of pulses, as a function of the tilt angle of the etalon.
(b) Tooth spacing of the high frequency comb (solid red line)
and the low frequency comb (ring cavity repetition rate - blue dashed line)
versus cavity perimeter.}
\label{group_velocities}
\end{figure}
Figure~\ref{group_velocities} shows the repetition rate dependence of ring laser as a function of the
angular tilt of the etalon with respect to the normal (a), and as a function of the cavity perimeter (b).
Figure (a) cannot be explained by
the angular dependence of the ``group delay'' in Eq.~(\ref{FPgroup_delay}).  Furthermore, there is a change
in repetition rate from 99.3072 MHz  (10.06976 ns round-trip time) to 99.01 MHz (10.098847) ns round-trip time)
by insertion of the Fabry-Perot, which is a change of round-trip time of 29.087 ps.
This difference does not corresponds to the insertion of the etalon, which should add $(n_g-1) d/c
= (1.462 -1)\times 15.119/c = 13.5667$ mm.  The measured 29.087 ps corresponds instead to the
insertion of an etalon of thickness 18.660 mm, or 3.54 mm more than the inserted glass!

Another evidence of the coupling between the modes of the laser and those of the etalon is the
plot of Fig.~\ref{group_velocities} (b) where the cavity length dependence of the
low frequency and high frequency combs are compared.  The pulse round-trip period in the etalon and the big ring are both linked to the
perimeter of the large ring cavity.  It should be noted that all these properties that have been
analyzed in reference~\cite{Masuda16} are only observed when the pulse duration
is much shorter than the Fabry-Perot etalon
round-trip time.

\subsection{Phase response}
\begin{figure} [h!]
\centering
\includegraphics*[width=\linewidth]
{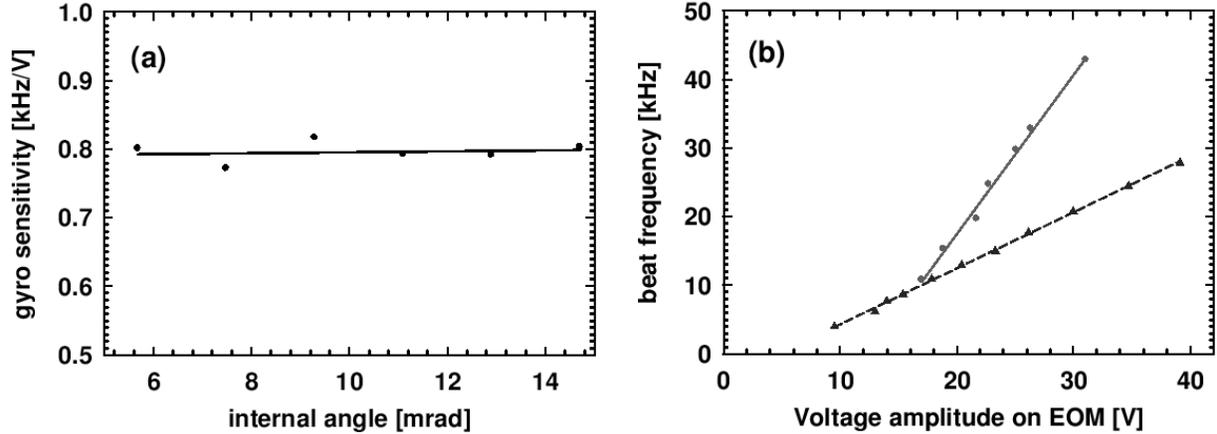} \caption[]{\small (a) Slope of the beat note response as a function of Fabry-Perot angle.
The slope remains at 0.8 kHz/V independently of the tilt of the Fabry-Perot.  (b) Comparison of the beat note response
before (red curve) and after (blue curve) insertion of the fabry-Perot. }
\label{group_vs_beat}
\end{figure}

It is clear from the previous section that the modes of the laser comb and Fabry-Perot are coupled,
and therefore it can be expected that the laser comb will be influenced by the dispersion of
the Fabry-Perot.  The previous measurements have also established that the
average pulse envelope velocity in the laser is {\em not related to $d\psi/d\Omega$} as is generally taken for granted.
Indeed, Figure~\ref{group_velocities} (a) has shown that the pulse envelope velocity varies significantly with
the angle $\theta$.  Over that range, $d\psi/d\Omega$ remains constant, and so is the phase response plotted
as a function of tilt angle in Fig.~\ref{group_vs_beat} (a).  In Fig.~\ref{group_vs_beat} (b), the
beat note is plotted as a function of applied voltage on the lithium niobate modulator, before (red solid curve) and
after insertion of the Fabry-Perot.   The gyro response slope switches from 2.3 kHz/V without Fabry-Perot, to 0.8 kHz/V, which implies, from Eq.~(\ref{basic_equation_disp}),
that:
\begin{equation}
1 +
\frac{1}{\tau_{\phi 0}}\left .\frac{d \psi}{d\Omega} \right |_\omega = 2.3/0.8 = 2.9
\label{FP_resonant_disp}
\end{equation}
therefore, from Eq.~(\ref{approx1}) we note that the ratio of the slopes minus one is
the product of the ratio of the laser to etalon optical lengths, times $(1+R)/(1-R)$:
\begin{equation}
\frac{nd}{c\tau_{\phi 0}}\frac{1 + R}{1 - R}  = 0.01456 \times \frac{1 + R}{1 - R} = 1.9.
\end{equation}
The ratio $(1+R)/(1-R)$ should thus be $1.9 \times 68.68 = 130$, which corresponds to a
value of effective reflectivity $R = 98$\%.  We note that this corresponds to the
reflectivity needed to create a bunch of 20 pulses.

\subsection{Negative versus positive dispersion}
In the case of an intracavity etalon, the modes of the laser couple to those of the etalon
because this is the configuration of minimum losses.
This is also the reason that
normal dispersion ($d \psi/d\Omega > 0$) is observed, as the Kramers-Kr\"onig correspondent of a
negative loss line.  Using the etalon in reflection would provide the negative dispersion
needed for amplification of the phase response, according to Eq.~(\ref{basic_equation_disp}).
However, the reflection characteristic would favor operation of the laser with the modes
{\em between} cavity resonances. The matching of the resonance could be forced by
active stabilization of the laser modes, or by synchronous pumping (OPO configuration),
which may also require active stabilization.  One solution that addresses the phase problem without
introducing
periodic losses is to substitute a cavity mirror by a Gires-Tournois interferometer.
The latter is essentially an etalon of which one face has 100\% reflectivity, and the other
face a field reflectivity $r$.
Its transfer function is given by~\cite{Diels06}:
\begin{equation}
{\cal R} =
\frac{-r+e^{-i\delta}}{1-re^{-i\delta}} = e^{-i \psi} \label{IF-2}
\end{equation}
where $\delta = 2 k d \cos \theta$ is the phase delay,
$\theta$ the {\em internal} angle. Near a resonance $\delta = 2 N \pi$,
the phase shift of the device can be approximated by:
\begin{equation}
\psi(\Omega) = - \left [\arctan (\frac{1+r}{1-r}) \right ] \delta
\end{equation}
Near the resonance, the group delay is approximately:
\begin{equation}
\frac{d \psi}{d\Omega} \approx - \left ( \frac{1+r}{1-r} \right ) \frac{d \delta}{d\Omega}.
\end{equation}
which has indeed the correct sign for enhancement of the gyro response.

Adding a Gires Tournois  of exactly the same thickness to the
present cavity will add a negative component to the denominator of Eq.~(\ref{basic_equation_disp}).
The Fabry-Perot with its resonances will lock the modes of the laser, which are then
also locked to those of the Gires-Tournois of the same thickness.  The magic
value of the reflectivity that will make the denominator of Eq.~(\ref{basic_equation_disp})
equal to zero is $r$ = 99\% (intensity reflectivity of $r^2 = 0.98$).

\section{Conclusion}

This work addresses the phase response of a sensor based on Intracavity Phase Interferometry.
The device is a mode-locked laser in which two pulses circulate in the cavity, and are given a
phase shift relative to each other by the physical quantity to be measured.
The response of the device is a
beat frequency between the two frequency combs issued from  the laser.  The beat frequency
is proportional to the phase shift (or physical parameter to be measured).
It is shown that the proportionality constant (between beat frequency and phase)
can be modified by introducing a giant dispersion {\em for each tooth} of  the frequency comb.
It is demonstrated experimentally with a mode-locked ring laser that the desired
coupling of a dispersion to all modes is obtained by inserting a low finesse etalon in
the laser cavity.  The beat note (or gyroscopic) response modification is
a large reduction (by a factor 2.9), because the Fabry-Perot etalon introduced a
positive (normal) resonant dispersion.  It is pointed out that a resonant negative dispersion,
that would enhance the beat note response, could be achieved with a
Gires-Tourois interferometer.  There are numerous other possibilities, involving for instance
two photon absorption, that can be exploited to achieve a resonant dispersion affecting
all modes of the frequency comb, to enhance the sensitivity of this class of devices.

\end{document}